# Quarkonia production at RHIC

**M. J. Leitch**
Los Alamos National Laboratory, P-25 MS H846, Los Alamos NM 87544, USA

E-mail: leitch@lanl.gov

**Abstract.** Quarkonia (J/$\psi$, $\psi'$, $\Upsilon$) production provides a sensitive probe of gluon distributions and their modification in nuclei; and is a leading probe of the hot-dense (deconfined) matter created in high-energy collisions of heavy ions. I will discuss our current understanding of the modification of gluon distributions in nuclei and other cold-nuclear-matter effects in the context of recent p-p and p(d)-A quarkonia measurements. Then I will review the latest results for nucleus-nucleus collisions from RHIC, and together with the baseline results from d-A and p-p collisions, discuss several alternative explanations for the observed suppressions and future prospects for distinguishing these different pictures.

## 1. Introduction

The physics and results for heavy-vector meson (quarkonia) production at RHIC in the context of similar results from lower energy fixed target measurements is reviewed. Besides this 1) introduction, this overview covers 2) production mechanisms in p+p collisions, 3) cold nuclear matter (CNM) effects for production in nuclei, 4) quarkonia results in the environment of the hot-dense matter created in A+A collisions, and then 5) concludes with comments about the future and a short summary.

The measurements and theoretical models shown here rely heavily on experimental colleagues at RHIC and elsewhere as well as on the insight gained by interactions over the years with many theoretical colleagues. Of special note are the talks on aspects of this subject given at this conference by Andrew Glenn, Johan Gonzalez, and Andre Rakotozafindrabe.

## 2. Quarkonia Production

Gluon fusion dominates the production of quarkonia, but the configuration of the initial $c\bar{c}$ state and how it hadronizes remains uncertain. Absolute cross sections can be reproduced by NRQCD models that produce a color octet state[1], but these models predict transverse polarization of the J/$\psi$ at large $p_T$ that is not seen in the data[2]. A general complication in understanding J/$\psi$ results is the fact that ~40% of the J/$\psi$s come from decays of higher mass resonances ($\psi'$ and $\chi_C$)[3] – a feature that may contribute to the lack of polarization seen, and also has important implications for CNM and QGP interpretations. One exception to this feature is the maximal transverse polarization observed for the 2S+3S $\Upsilon$ states[4] (Figure 1); presumably the feed-down mentioned above might help destroy any polarization for the J/$\psi$ and $\Upsilon_{1S}$, but allow polarization to persist for the $\Upsilon_{2S+3S}$ where there is no feed-down. Accurate measurements of the $\psi'$, where feed-down is also absent, would help to clarify this issue. The existing measurements from CDF[5] lack the precision to address this.

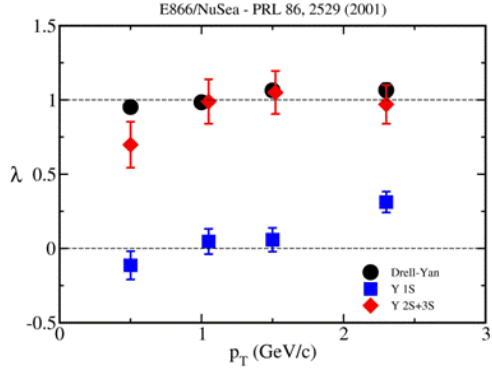

Figure 1 – Polarization ($\lambda$) of $\Upsilon$ and Drell-Yan production from E866/NuSea for 800 GeV fixed target measurements[4].

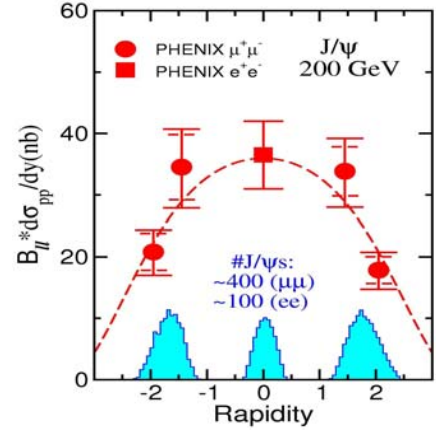

Figure 2 - J/$\psi$ differential cross section vs. rapidity in 200 GeV p+p collisions from the 2003 PHENIX data[6].

J/$\psi$ cross section measurements for p+p at $\sqrt{s}$=200 GeV from PHENIX[6] are shown in Figure 2. These results, based on approximately 500 J/$\psi$s from the 2003 run, provide the baseline for both CNM studies in d+Au collisions and QGP studies in A+A collisions at RHIC, and are presently one of the limiting factors in obtaining precise nuclear modifications. However p+p data from the 2005 and 2006 runs will soon improve this baseline significantly with over 40,000 J/$\psi$s expected from analysis of this newer data.

## 3. Cold Nuclear Matter (CNM) effects on Quarkonia

When quarkonia are produced in nuclei their yields per nucleon-nucleon collision are known to be significantly modified. This modification, shown vs. $x_F$ in Figure 3 for 800 GeV p+A fixed target measurements by E866/NuSea, is thought to be due to several CNM effects including gluon shadowing, initial-state gluon energy loss and multiple scattering, and absorption (or dissociation) of the $c\bar{c}$ in the final-state before it can form a J/$\psi$.

Shadowing is the depletion of low-momentum partons (gluons in this case) in a nucleon embedded in a nucleus compared to their population in a free nucleon. The strength of the depletion differs between numerous models by up to a factor of three. Some models are based on phenomenological fits to deep-inelastic scattering and Drell-Yan data[7], while others obtain shadowing from coherence effects in the nuclear medium[12,8]. The gluon shadowing from the Frankfurt and Strikman[12] is shown in Figure 4. In addition, models such as the Color Glass Condensate (CGC)[9] also yield shadowing through gluon saturation pictures where the large gluon populations at very small x in a nucleus generate a deficit of gluons at small x via two-to-one gluon diagrams. This effect exists even in a free nucleon, but is amplified in a nucleus by a factor of $A^{1/3}$ where it then produces nuclear shadowing.

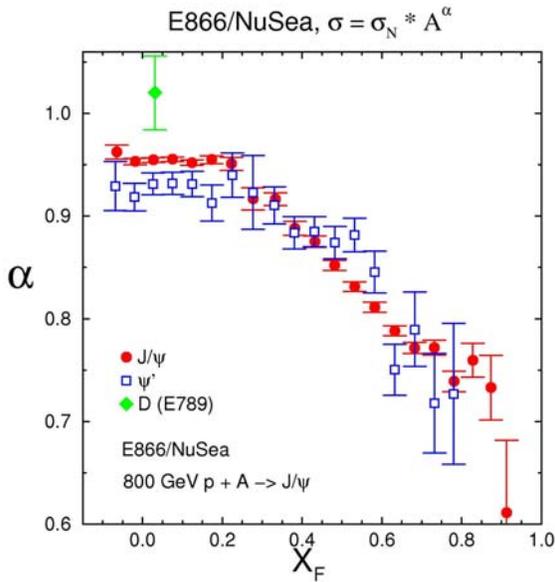
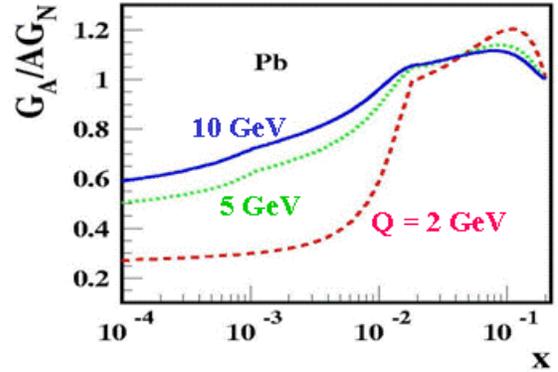

Figure 4 - Gluon shadowing for Pb from Ref 12 showing depletion of the gluon structure function in a nucleus at small x.

Figure 3 - Nuclear modification factor $\alpha$ vs. $x_F$ for J/$\psi$ and $\psi'$ from E866/NuSea[10] and $D^0$ from E789[11]. $\alpha$ is defined as shown on the top of the plot.

In the final state, the produced $c\bar{c}$ can be disassociated or absorbed, as shown in Figure 8, on either the nucleus itself, or on light co-moving partons produced when the projectile proton or deuteron enters the nucleus. The latter is probably only important in nucleus-nucleus collisions as the number of co-movers created in a p+A or d+A collisions is small compared to those produced in A+A collisions.

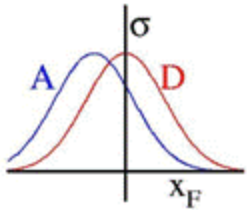
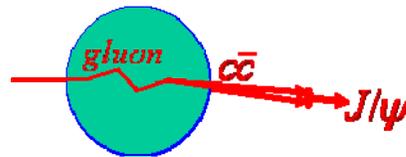

Figure 5 - Cartoon illustrating the shift vs. $x_F$ of the differential cross section due to energy loss of a gluon in the initial state.

Figure 6 - Cartoon illustrating initial-state gluon multiple scattering that broadens the $p_T$ of J/$\psi$s produced in a nucleus.

Another theoretical model from Brodsky[13] asserts that many of the J/$\psi$s that are produced come from intrinsic heavy-quark fock states in the incident proton wave function, especially for large J/$\psi$ $x_F$. In this case an $A^{2/3}$ suppression results from the stripping off on the nuclear surface of the light-quark components in order to free the heavy-quark fluctuation and allow it to hadronize as shown in
Figure 9.

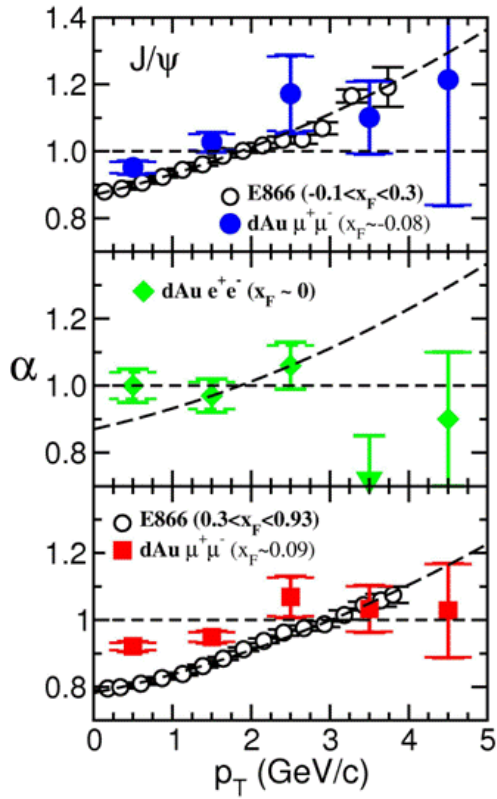

Figure 7 - $p_T$ broadening in three $x_F$ ranges for J/ψ production at 200 GeV[6] (colored points) compared to that from fixed-target measurements at 39 GeV[10] (open black points).

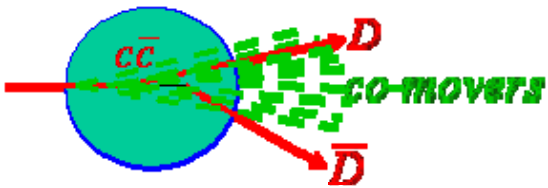

Figure 8 – Cartoon illustrating final-state absorption (or disassociation) of $c\bar{c}$'s in the nucleus or by co-movers.

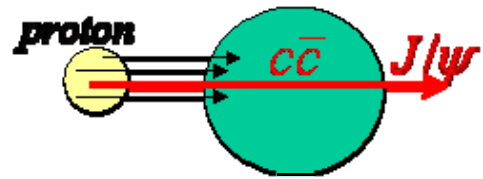

Figure 9 – Cartoon illustrating intrinsic charm contribution to J/ψ production

However, J/ψ suppression in p(d)+A collisions remains a puzzle given that one does not find a universal suppression vs $x_2$ as would be expected from shadowing, (Figure 10a); while vs. $x_F$ the dependence is more similar for all energies (Figure 10b), especially for the lower energy measurements where a large range in $x_F$ is covered. This apparent $x_F$ scaling supports explanations that involve initial-state energy loss or Sudakov suppression[14].

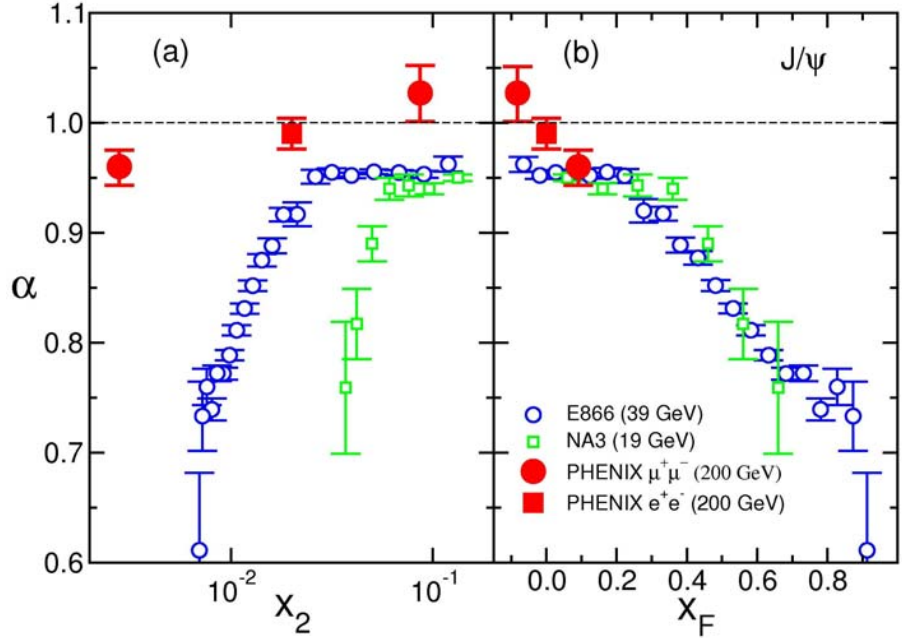

**Figure 10** - Test of scaling vs. $x_2$ and $x_F$ for J/ψ suppression data for three different collisions energies. Data is from Refs. 6, 10 and 15.

## 4. Quarkonia in the Hot-dense matter from A-A Collisions

One of the leading predictions for the hot-dense matter created in high-energy heavy-ion collisions was that if a deconfined state of quarks and gluons was created, i.e. a quark-gluon plasma (QGP), the heavy-quark bound states would be screened by the deconfined colored medium and destroyed before they could be formed[16]. This screening would depend on the particular heavy-quark state, with the ψ' and $\chi_C$ being dissolved first; next the J/ψ and then the ϒs only at the highest QGP temperatures. The CERN SPS measurements showed a suppression for the J/ψ and ψ' beyond what was expected from CNM effects - as represented by a simple absorption model constrained to p+A data. In addition to explanations involving creation of a QGP, a few theoretical models[17] were also able to explain the data[18] without including a QGP, so the evidence that a QGP was formed was controversial.

The first measurements from PHENIX at RHIC in 2004 are beginning to yield results – see Figure 11 for preliminary results for Au+Au and Cu+Cu collisions[19]. First it is critical to understand what the normal or CNM J/ψ suppression should look like in these A+A collisions. This is illustrated by the blue error bands for A+A collisions in Figure 11 which represent identical theoretical calculations[20] to the analogous blue error band in Figure 12 for d+Au collisions. As can be seen the d+Au data lacks enough precision (because of the low statistics obtained so far in d+Au) to provide a good constraint on the CNM effects. As a result the uncertainty of the CNM effects in A+A collisions makes it difficult to be very quantitative about the amount of "anomalous" suppression observed, although there does seem to be a clear suppression beyond CNM for the most central collisions (right-most points on the A+A plots).

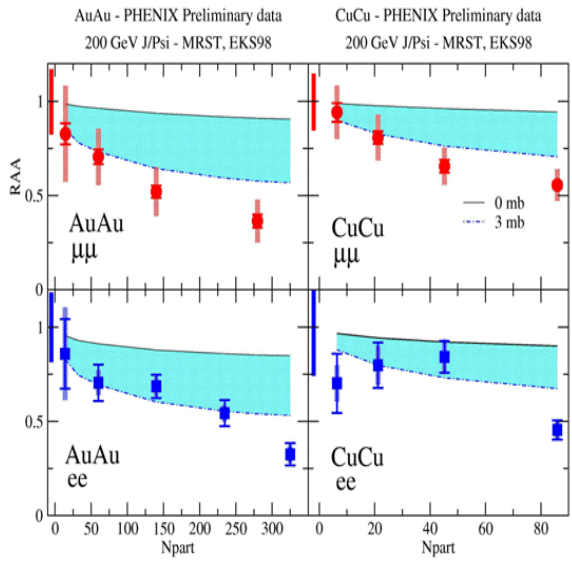
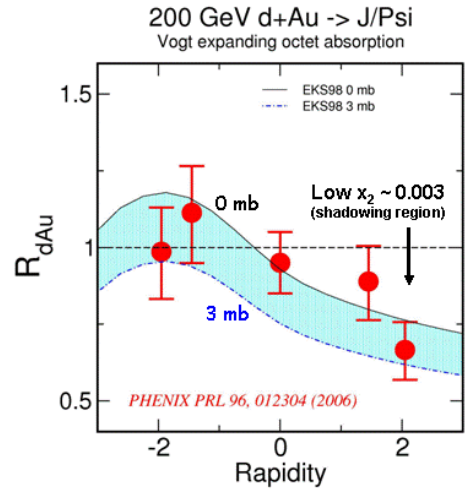

**Figure 11 - J/ψ suppression in Au+Au and Cu+Cu collisions (PHENIX preliminary[19]) for forward rapidity (μμ) and central rapidity (ee) compared to predictions for CNM from the same calculations as shown in Figure 12.**

**Figure 12 - Results for J/ψ suppression in d+Au collisions[6] compared to a theoretical calculation[20] that includes absorption and EKS shadowing.**

On the other hand, all of the models[17,21,22] that were successful in describing the lower energy SPS data over-predict the suppression compared to the preliminary data at RHIC - unless a "regeneration" mechanism is added as was done by Rapp[22] and by Thews[23]. The regeneration models assert that if the total production of charm is high enough then densities in the final state will be sufficient to have substantial formation of J/ψs from the large number of charm quarks created in the collision. This production mechanism was almost insignificant at SPS energies but at RHIC may be substantial. This leads to the scenario for J/ψ suppression at RHIC in which strong screening or dissociation by a very high-density gluon density is occurring to a level of suppression stronger than the data shows, but the regeneration mechanism compensates for this and brings the net suppression back up to where the data lies. This "coincidence" is somewhat unsatisfying, but it should be noted that Rapp's calculations were true predictions made before the experimental results and do match the data quite well – at least as far as the centrality dependence shown in Figure 13.

An alternative interpretation of the preliminary results, sequential screening, is given by Karsch, Kharzeev and Satz[24]. In this picture, they assume that the J/ψ is never screened, as supported by recent Lattice QCD calculations for the J/ψ [25] – not at SPS nor at RHIC. Then the observed suppression comes from screening of the higher-mass states alone (ψ' and $\chi_C$) that, by their decay, normally provide ~40% of the observed J/ψs. This scenario is then consistent with the apparently identical suppression patterns seen at the SPS and RHIC shown in Figure 14.

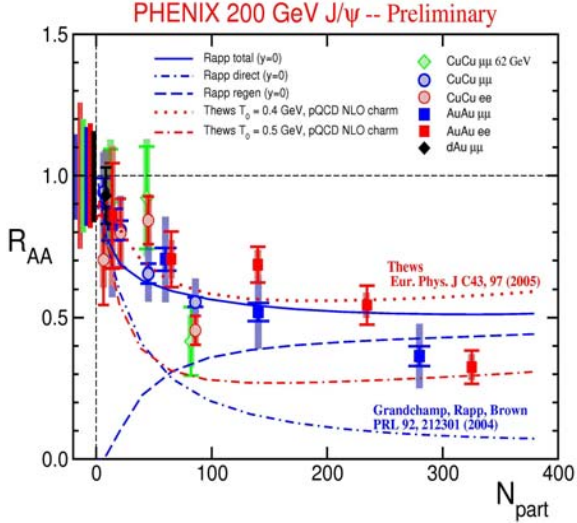 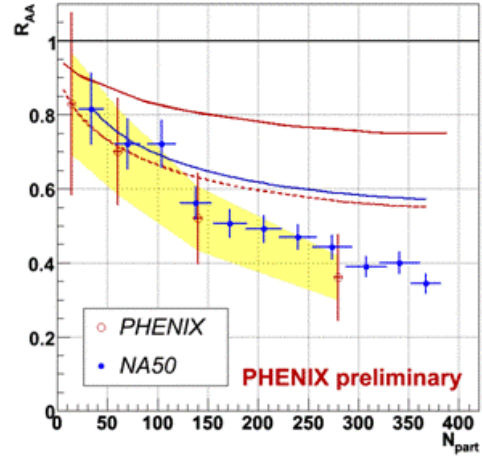

**Figure 13** - Two of the regeneration models that can explain the J/ψ suppression observed at RHIC. All data is at 200 GeV except for the 62 GeV points indicated. The solid curve is the full calculation from Ref. 22. The direct (dot-dash) and regeneration (dashed) alone curves are also shown. The dotted curve is from Ref. 23, with another version for larger temperature also shown (dot-dash-dash).

**Figure 14** - The sequential screening picture where SPS and RHIC data follow a universal behavior vs. number of participants. The curves are absorption calculations for SPS with $\sigma_{abs}$ = 4.18 mb (blue); and 1 mb (solid), 3 mb dashed at RHIC (red).

As a result we are left for the moment with two different scenarios that provide explanations for the RHIC A+A data. Both include the QGP in their picture, either through color screening in the QGP or through severe suppression of the J/ψ by a very high gluon density. Further tests from the data will be necessary to clarify the picture. Regeneration models predict narrowing of both the rapidity and $p_T$ distributions, but so far the preliminary data shows little or no change in the rapidity shape from ordinary p+p and only a hint of narrowing of the $p_T$. We are also trying to extract a measurement of flow for the J/ψ, since emerging results for single charm are beginning to show flow[26] and the J/ψ's, if they were from regeneration, would inherit this flow. These tests await more precise final analysis of the 2004,5 Au+Au and Cu+Cu data; and higher statistics runs for Au+Au and d+Au in the near future.

## 5. Future and Summary

As RHIC luminosities increase, the precision of studies of the rare quarkonia probes will improve and more quantitative analysis will be enabled. Already PHENIX has observed the first ϒs in the 2005 run (Figure 15) and also J/ψs in Ultra-peripheral collisions (Figure 16). In addition the first J/ψs have been observed by STAR in their 2005 data, as shown by Gonzalez at this conference.

Substantial uncertainties in the production mechanism remain that propagate to further uncertainty in interpretations of p(d)+A and A+A data. In d+Au collisions at RHIC, weak suppression is seen, but comparisons between data at different energies suggests that shadowing is not the relevant physics for the observed suppression and a clean understanding of the CNM effects eludes us. Finally, the initial results from RHIC for A+A collisions support the creation of hot-dense matter that destroys J/ψs either through screening in a deconfined medium or disassociation by very large gluon densities. An understanding of whether models with regeneration or sequential screening are the correct description of the A+A results awaits more precise data.

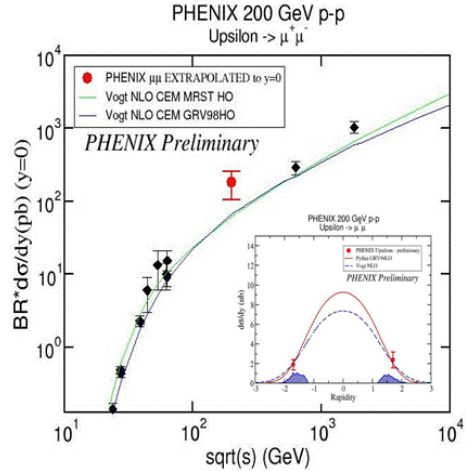

**Figure 15 – Cross section (red point at 200 GeV) for the first ϒs at RHIC observed via muon pairs at large rapidity (red points on inset) in PHENIX.**

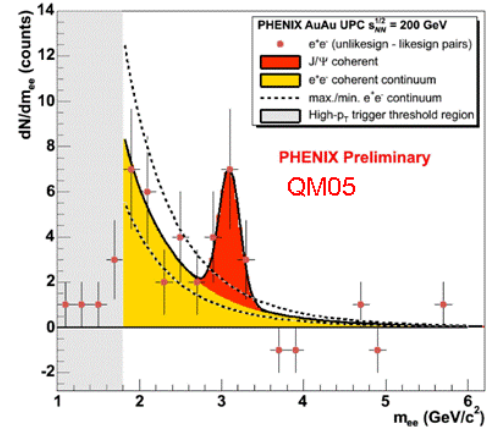

**Figure 16 - First J/ψs observed by PHENIX in ultra-peripheral Au+Au collisions**